\documentclass[10pt,a4paper]{article}
\usepackage{natbib}
\usepackage[english]{babel}
\usepackage{amsmath,amsthm,amsbsy,amssymb,amsfonts}
\usepackage{lineno}
\usepackage{mathtools}
\usepackage{graphicx}
\usepackage{epstopdf}
\usepackage{multirow}
\usepackage{url}
\usepackage{bigints}
\usepackage{xcolor}
\newtheorem{theorem}{Theorem}

\newcommand*{\myDots}{\ifmmode\mathellipsis\else.\kern-0.13em.\kern-0.13em.\fi} % touching at \kern-0.1725em

\def\ept{\mathop{\bf E}\nolimits}
\newcommand{\ee}{{\rm e}}
\newcommand{\dd}{{\rm d}}

\newcommand{\hdoublespacing}{\baselineskip=21pt plus 2.5pt minus 1pt}

\RequirePackage[normalem]{ulem} %DIF PREAMBLE
\RequirePackage{color}\definecolor{RED}{rgb}{1,0,0}\definecolor{BLUE}{rgb}{0,0,1} %DIF PREAMBLE
\providecommand{\DIFadd}[1]{{\protect\color{blue}\uwave{#1}}} %DIF PREAMBLE
\providecommand{\DIFdel}[1]{{\protect\color{red}\sout{#1}}}                      %DIF PREAMBLE
\newcommand{\remark}[1]{\textcolor{orange}{\it \small [#1]}}

\begin{document}
%\linenumbers

\title{\large\bf How to quantify  earthquake predictability? \\  Advances in earthquake forecasting and predictability limits
%: A theoretical approach
}
\author{\it\small Jiancang Zhuang$^{1,2,3}$ \&   Didier Sornette$^{3}$ \\
\footnotesize{\it $^1$The Institute of Statistical Mathematics, Research Organizatoin of Information and Systems}\\ \footnotesize{\it 10-3 Midori-cho, Tachikawa, Tokyo 190-8562, Japan}\\
\footnotesize{\it $^2$Department of Statistical Science, the Graduate University for Adavanced Studies (SOKENDai)}\\ \footnotesize{\it 10-3 Midori-cho, Tachikawa, Tokyo 190-8562, Japan}\\
\footnotesize{\it $^3$Institute of Risk Analysis, Prediction and Management,  Southern University of Science and Technology}\\
\footnotesize{\it 1088 Xueyuan Ave., Nanshan District, Shenzhen 518055, China}}
%\date{\today}
\maketitle

\hdoublespacing

\section*{Abstract}

Earthquakes resist deterministic prediction, yet their occurrence is not fully random. This paper develops a unified information-theoretic framework to quantify predictability. By reviewing Shannon entropy and the Kullback--Leibler divergence, we formalize predictability as the entropy gap between complete randomness and the true data-generating process and clarify how this absolute notion relates to the relative skill gains used in prospective model evaluation.
Within the point-process setting, we derive entropy rates for the Poisson process and for ETAS and identify the intrinsic predictability rate as an information gain functional of the conditional intensity. Using this lens, we summarize what is currently established about earthquake predictability in time, space, and magnitude: temporal and spatial predictability are dominated by clustering and heterogeneous background rates, while magnitude predictability requires separating marginal magnitude statistics (e.g., Gutenberg--Richter and tapered laws) from genuine inter-event dependence encoded by the multivariate magnitude distribution.
Finally, we show how incorporating high-dimensional pre-event observations can increase predictability through mutual information, thereby reframing forecasting progress as the extraction of structured dependence between available information and future seismicity. This perspective provides a coherent basis for assessing predictability limits, comparing models, and identifying where additional information and physics that are most likely to yield substantive forecasting improvements.

\section{Introduction}

{The question of whether reliable prediction of individual earthquakes is scientifically achievable has been debated for more than half a century. By the late 1990s, the debate had reached a turning point.} \citet{geller1997science} {argued that reliable prediction of individual earthquakes was unlikely to be achievable in any practically useful sense. In contrast,} \citet{wyss1997science} {emphasized that this conclusion should not be interpreted as implying that earthquake occurrence is completely unpredictable. Instead, he argued that the earthquake generation process itself may still possess measurable predictability and that this predictability should be investigated scientifically.}

{This shift in perspective was articulated collectively during the seven-week \textit{Nature} debate, ``Is the reliable prediction of individual earthquakes a realistic scientific goal?'' \citep{nature1999earthquakedebate,nature1999earthquakedebate_continuation}. The discussion brought together sharply contrasting views on both the feasibility and the proper scientific meaning of earthquake prediction. Some contributors argued that decades of unsuccessful searches for reproducible precursors cast serious doubt on deterministic predictions specifying the time, location, and magnitude of individual earthquakes. Others considered claims of intrinsic unpredictability premature, emphasizing the limited observations, incomplete understanding of rupture nucleation, and insufficient investment in multidisciplinary research. Intermediate positions sought to determine the boundary between what is predictable and what is not. They stressed that earthquakes are neither periodic nor completely random: seismicity exhibits fault localization, aftershock triggering, temporal clustering, scale invariance, and possible collective or critical organization. The debate therefore shifted attention from binary claims that earthquakes either can or cannot be predicted towards the quantitative assessment of predictability, rigorous prospective testing, probabilistic forecasts of seismic rates and hazards, and the search for physical, statistical, and multidisciplinary constraints capable of improving those forecasts. This conceptual evolution laid the foundation for later developments in operational earthquake forecasting and quantitative studies of intrinsic earthquake predictability.}

Back to the early 1970s,  a major line of researches  associated with Gelfand, Keilis-Borok, Knopoff, Press, Rikitake, Vere-Jones, and others, developed pattern-recognition algorithms, morphostructural criteria, stochastic point-process models, and formal statistical tests of prediction schemes. Landmark contributions such as \cite{gelfand1972tectonophysics,gelfand1976pepi} explicitly sought to identify systematic precursory structures in seismicity and to evaluate their predictive performance. These and subsequent works introduced concepts such as ``times of increased probability,'' long-term seismic activation patterns, renewal processes, and likelihood-based model comparison. In parallel, substantial work had already framed earthquake prediction within formal probabilistic and statistical paradigms. Building on conceptual foundations such as \citet{aki1989tectonophysics}, studies including \citet{dvj1995ijf,dvj1998compseis} formulated prediction as the estimation of conditional probabilities for future seismicity, thereby distinguishing deterministic prediction from probabilistic forecasting.

In the 2000s, these discussions were sharpened conceptually. \citet{jordan2006srl} articulated a clear operational distinction between scientific prediction, in the form of a testable hypothesis about the time, location, and magnitude distribution of future events, and useful prediction, a forecast sufficiently accurate and timely to support mitigation actions. He further emphasized the idea of defining intrinsic predictability as the degree to which the future evolution of seismicity is determined or constrained by the prior state and history of the fault system, independent of observational limitations. {Although closely related ideas had appeared earlier in statistical seismology and pattern-recognition research, t}his clarified the distinction between limits imposed by the physics of the system and those arising from incomplete models or data, and helped motivate the development of prospective, community-based testing frameworks such as the Collaboratory for the Study of Earthquake Predictability (CSEP) and pragmatic approaches for providing near real-time probabilistic forecasts of seismicity such as operational earthquake forecasting (OEF) \citep{jordan2011ag,mizrahi2024rg}.

In much of this literature, however, predictability is treated largely as an intuitive or qualitative notion: models are compared and tested, but the fundamental degree to which seismicity is predictable is seldom defined as a measurable quantity.
In fact, predictability can be defined and quantified rigorously from the mathematical standpoint in terms of information theory as the reduction of uncertainty about future seismicity provided by knowledge of past observations and model structure. In this article, we develop such an information-theoretic formulation of earthquake predictability, summarize what has been learned so far about earthquake predictability, and discuss how to improve earthquake forecasting.  Within this framework, intrinsic predictability corresponds to the maximum achievable information gain about the future given the true underlying process, whereas practical predictability (forecasting performance) reflects the information captured by specific forecasting models.

\section{Predictability}
\subsection{How to quantify information}
 Let us start with a simple example. Suppose we are tossing two coins, A and B. For Coin A, the probability of heads is $p_{_\mathrm{A}}$, and the probability of tails is $1-p_{_\mathrm{A}}$. Similarly, the probabilities for Coin B are $p_{_\mathrm{B}}$ for heads and $1-p_{_\mathrm{B}}$ for tails. We assume that the outcomes of the two coin tosses are independent.
Now, suppose that we need a function or operator that provides the uncertainty contained in a stochastic system by acting on the sets of probabilities for each state in the system. The requirement for such an information function or operator, namely $f$, is that observing two uncorrelated coins independently should yield the same amount of uncertainty as observing them synchronously.
That is,
\begin{equation}\label{eq:twocoin}
  f(\{p_{_\mathrm{A}} , 1-p_{_\mathrm{A}}\})+f(\{p_{_\mathrm{B}}, 1-p_{_\mathrm{B}}\})= f(\{p_{_\mathrm{A}}p_{_\mathrm{B}}, (1-p_{_\mathrm{A}})p_{_\mathrm{B}}, (1-p_{_\mathrm{A}})p_{_\mathrm{B}}, (1-p_{_\mathrm{A}}) (1-p_{_\mathrm{B}})\})
\end{equation}

{For a general case of a system with $n$ possible outcomes, a} natural candidate for $f$
that satisfies the above property can be written as
\begin{equation}
  f(\{p_1, p_2, \cdots, p_n\})=-\sum_{i=1}^n p_i\log_c p_i, ~~~\mbox{with }~~~ \sum_{i=1}^n p_i=1,
\end{equation}
where {$p_i, \, i=1,\, 2, \cdots, n$, denotes the  probability of the $i$th outcome,} $c$ is any positive number, typically taking any positive value, such as $\ee$ or $2$.
Note that this form is not the unique solution if one assumes only independent additivity. Its uniqueness additionally requires a further condition such as the chain rule or an equivalent grouping/recursivity axiom (see, e.g., \citealp{shannon1948info,cover2005eoit}).

In fact, the above form defines the Shannon information entropy (\citealp{shannon1948info}). For a discrete system,  the Shannon information entropy is defined by
\begin{equation}
  H(X):=-\sum_{x\in \Omega} p_X(x) \log p_X(x)
\end{equation}
and for a continuous system, it is defined as
\begin{equation}
  H(X):=-\int_\Omega  p_X(x) \log p_X(x)\,\dd x,
\end{equation}
where $\Omega$ is domain of the possible values of $X$, $p_X(x)$ is defined via $p_X(x)=\Pr\{X=x\}$ for discrete case and $p_X(x)\,\dd x=\Pr\{X\in (x,x+\dd x)\}$ for continuous case.

The meaning of the Shannon entropy is not so easy to grasp. However, if we choose 2 as the base of the logarithm to replace the natural logarithm in the definition, the Shannon entropy for the symmetric Bernoulli trial $X$ ($p=1/2$),
\begin{equation}
  H(X)=-[0.5 \log_2 0.5 +(1-0.5)\log_2 (1-0.5)]=1,
\end{equation}
implying that the uncertainty carried by a symmetric Bernoulli trial is the same as that carried by a binary digit (bit) in our computer system. Therefore, bit can naturally serve as the unit of uncertainty. Figure~\ref{fig:one}(a) shows how the system entropy varies as a function of $p$. The entropy attains its maximum at $p = 1/2$, corresponding to the case where both outcomes are equally likely. At this point, the system exhibits maximal uncertainty: no outcome is favored, and the next state cannot be anticipated better than random guessing. In this sense, the system is least predictable, as no information about past behavior provides a reduction in uncertainty about the future.

\newcommand{\figone}{(a) Entropy of different Bernoulli trials characterized by varying toss probability $p$. (b) Expected negative log-likelihood for forecasting a Bernoulli trial system with $p=1/4$ (true value) as a function of the model forecasting probability $p^*$. (c) Filled contour plot of the expected negative log-likelihood for the Bernoulli trial system as a function of the true probability $p$ and the forecast probability $p^*$.  (d) Comparison between the entropy of the Bernoulli trial system (red curve) and the expected negative log-likelihood under varying forecast probabilities $p^*$ for a true system with $p=1/4$ (black curve).}
%Panel (a) shows the entropy of Bernoulli trials as a function of the success probability $p$.
%Panel (b) displays the expected negative log-likelihood for a true system with $p=1/4$, plotted as a function of the forecast probability $p^*$.
%Panel (c) compares the true entropy (red curve) with the expected log-loss under varying forecast models (black curve).}
\begin{figure}
\begin{tabular}{cc}
\bf \large (a) & \bf\large (b) \\
{\includegraphics[trim=0 0 1724 80,clip,width=0.425\textwidth]{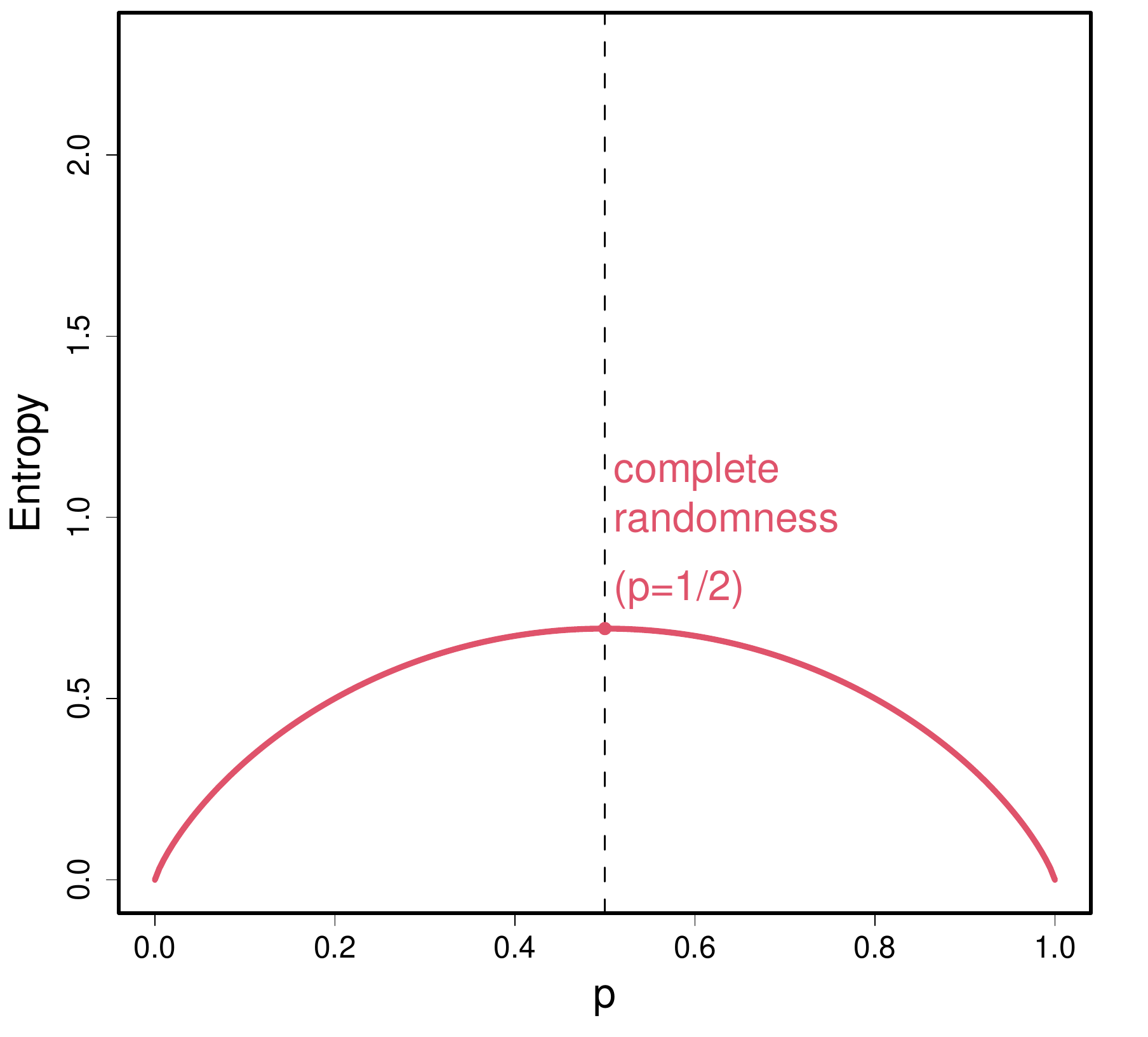}}&
{\includegraphics[trim=862 0 862 80,clip,width=0.425\textwidth]{fig1.pdf}}\\
\bf\large (c)& \bf\large (d)\\
{\includegraphics[trim=0 180 30 195,clip,width=0.455\textwidth]{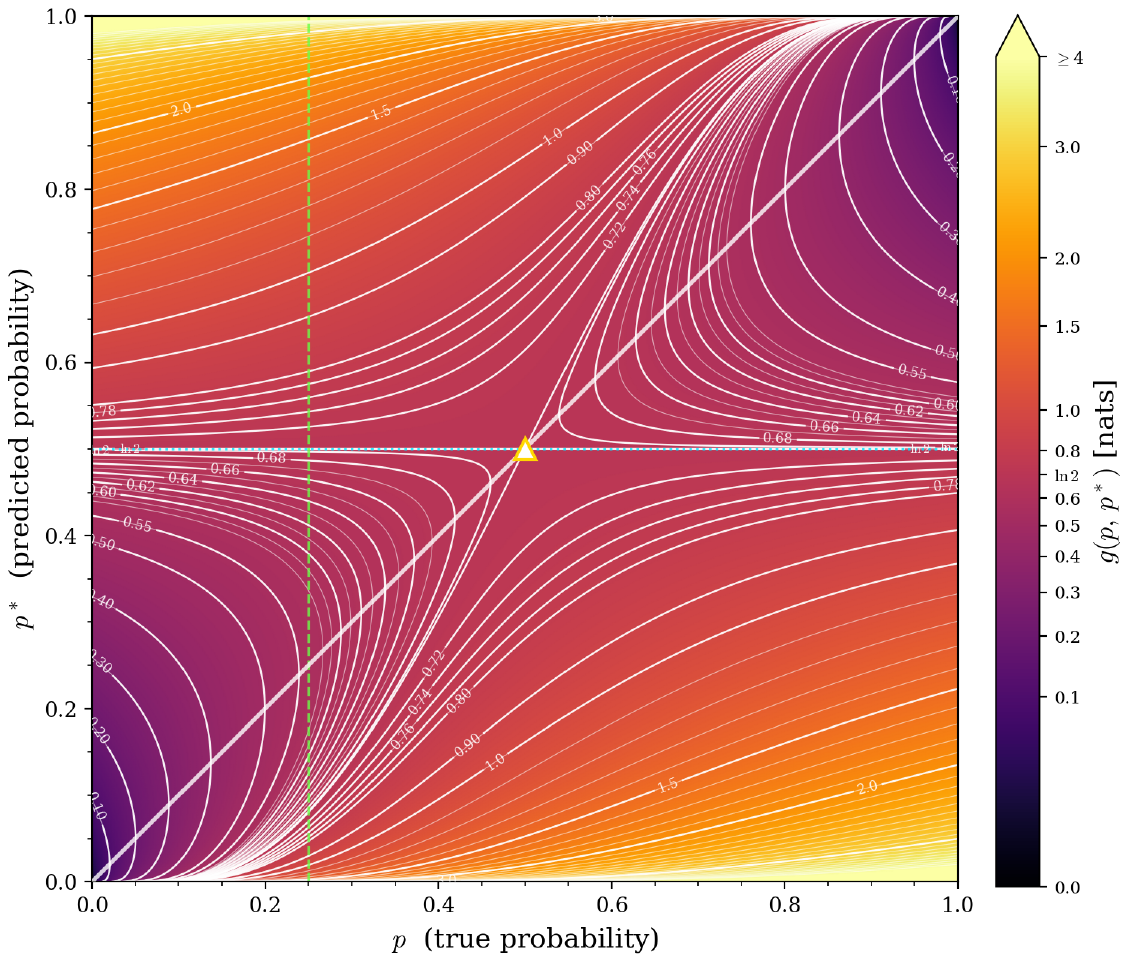}}&
{\includegraphics[trim=1724 0 0 80,clip,width=0.425\textwidth]{fig1.pdf}}
\end{tabular}
\caption{\label{fig:one}\figone}
\end{figure}
\subsection{How to quantify forecasts}
Consider again the Bernoulli random variable $X$. Suppose a forecasting model assigns probability
$p^* := \Pr\{X=1 \mid \text{forecasting model}\}$,
while the true data-generating probability is
$p := \Pr\{X=1 \mid \text{true model}\}$.
The expected negative log-likelihood (log-loss) of this forecast, evaluated with respect to the true distribution, is given by
\begin{eqnarray}
g(\{p,1-p\}, {p^*,1-p^*}) &:=& -\ept[I(X=1)\log p^* + I(X=0)\log (1-p^*) ]\nonumber\\
&=&-p \log p^* - (1-p)\log (1-p^*),
\end{eqnarray}
where $I(\cdot)$ is the indicator function. The above expression can be interpreted as a measure of discrepancy between the forecasting model $p^*$ and the true model $p$ (see Figures~\ref{fig:one}(b) and (c)).
More precisely, it corresponds to the cross-entropy between the true distribution and the forecast.
This discrepancy is minimized if and only if $p^* = p$, in which case the log-loss reduces to the Shannon entropy of the true distribution.
For any $p^* \neq p$, the excess loss is strictly positive, reflecting the additional uncertainty introduced by model misspecification.

%\begin{figure}
%\centerline{\includegraphics[trim=600 117 30 120,clip,width=0.42\textwidth]{mcs2/slide6.jpg}}
%\caption{\label{fig:bern.fcast}Expected negative likelihood for forecasting a Bernoulli trial systems with $p=1/4$.}
%\end{figure}

The above expected negative likelihood is called the
Kullback-Leibler divergence (entropy distance)
\begin{equation}
  D_{\mathrm{KL}}(P\|Q) := \left\{\begin{array}{cl}
   \displaystyle \sum_{x\in \mathbb{X}} p(x)\log \frac{p(x)}{q(x)},&\mbox{system with discrete outcomes,}\\\displaystyle \int_{ \mathbb{X}} p(x)\log \frac{p(x)}{q(x)}\,\dd x ,&\mbox{system with continuous outcomes,}
    \end{array}
    \right.
\end{equation}
where $\mathbb{X}$ is the set of values that the random variable $X$ can take.
Here \(P\) (with probability mass function or probability density function \(p(x)\)) is typically regarded as the true distribution, and \(Q\) (with \(q(x)\)) as the candidate (testing or forecasting) distribution.
The first step in forecasting is to find the distribution $Q$ which minimizes the observation on the Kullback-Leibler divergence. Since \(P\) is unknown, \(D_{\mathrm{KL}}(P\|Q)\) cannot be computed directly. In practice, we instead minimize an empirical counterpart by replacing the expectation under \(P\) with a sample average (equivalently, by using the empirical distribution of the observed samples). This is equivalent to maximizing the (log-)likelihood of the observations \(X\) under \(Q\), i.e., performing maximum likelihood estimation.

\subsection{Complete randomness and predictability}

Given a specified class of probability distributions, the one with the maximum entropy (not less than that of any other member) is called the {\em maximum entropy probability distribution (MEPD)}. According to the principle of maximum entropy, when only partial information about a distribution is available, typically expressed in the form of constraints defining an admissible class, the maximum entropy probability distribution (MEPD) should be selected as the least informative default. In this sense, it represents a model of complete randomness consistent with the imposed constraints.
Equivalently, the maximum entropy distribution is the one that introduces the least additional structure beyond the specified properties. It incorporates exactly the information contained in the constraints, and nothing more.

For example, common one-dimensional maximum-entropy distributions include the symmetric Bernoulli trial on two outcomes (zero and one), the geometric distribution on the positive integers with a fixed mean, the uniform distribution on a fixed finite interval, the exponential distribution on the nonnegative real line with a fixed mean, and the Gaussian distribution on the entire real line with a fixed mean and a fixed variance.

It is natural to quantify the predictability of a system as the reduction in uncertainty relative to a corresponding system exhibiting complete randomness.
Formally, we define predictability as the difference between the Shannon entropy of the maximum-entropy (fully random) system and the entropy of the true system.

For a Bernoulli variable, complete randomness corresponds to the fair coin with $p=1/2$, which maximizes entropy.
Consider a coin $A$ with head probability $p=0.25$ and a fair coin $B$ with $p=0.5$.
Since $H(\cdot)$ is maximized at $p=1/2$, coin $A$ is more predictable than coin $B$.
Its predictability can be quantified as
\begin{equation}
H(B) - H(A)
= \left[-0.5 \log 0.5 - 0.5 \log 0.5\right]
  - \left[-0.75 \log 0.75 - 0.25 \log 0.25\right]
= 0.75 \log 3 - \log 2
\approx 0.131,
\end{equation}
where logarithms are taken in base $e$ (or base 2 if predictability is measured in bits).

Figure~\ref{fig:one}(d) illustrates the relationship between the Shannon entropy of the true system and the expected negative log-likelihood (log-loss) obtained when different models are used to forecast the outcomes of coin $A$.
The minimum of the black curve occurs at $p^* = p = 1/4$, corresponding to the maximum-likelihood (and information-theoretically optimal) forecast.
For $p^* \neq p$, the expected log-loss exceeds the true entropy by an amount equal to the Kullback-Leibler divergence $D_{\mathrm{KL}}(p \| p^*)$.

\section{Predictability in Point Processes and in the ETAS Model}

\paragraph{Entropy rate of the Poisson process} For a temporal point process observed in a time interval $[0, \,T]$ with an expected rate $\lambda_0$, the process corresponding to complete randomness is the Poisson process with the same rate of $\lambda_0$. In this case, the likelihood is
\begin{equation}
  \log L = N \log \lambda_0 -\lambda_0 T,
\end{equation}
where $N$ is the number of events occurring in $[0,\, T]$. Given that $\ept [N]=\lambda_0 T$,  the entropy rate of the process is
\begin{equation}
  \epsilon  =\frac{1}{ T}{\ept [-\log L]} = \lambda_0 \left[ 1- \log \lambda_0\right].
\end{equation}

\paragraph{Entropy rate of the ETAS model} Let us consider the ETAS model (\citealp{ogata1988jasa}). This model is a marked  point process model and is completely determined by an intensity function, which  represents the rate (or likelihood) of an earthquake of magnitude $m$ occurring at time $t$ given the history of previous earthquakes.
In its temporal-only version, it takes the form
\begin{equation}\label{eq:etas}
\lambda(t,m)=\lambda_g(t)\, s(m) ~~~\mbox{with}~~~~ \lambda_g(t) :=\mu+\sum_{i:\, t_i<t} \kappa(m_i)\, g(t-t_i).
\end{equation}
In this above equation:
\begin{description}
\item (a) \(\mu\) is the background seismicity rate, accounting for the baseline occurrence of earthquakes that are independent of prior seismic activity, such as those caused by tectonic stress;
\item (b) The summation term \(\sum_{i:\, t_i < t} \kappa(m_i)\, g(t - t_i)\)  represents the contributions of past earthquakes to the current time, where
   the productivity function
   \begin{equation}
   \kappa(m)=A\,\exp(\alpha (m-m_0))
   \label{wh2ybgq}
   \end{equation}
     quantifies the aftershock-generating potential of an earthquake based on its magnitude \(m_i\), {$m_0$ denotes the magnitude threshold of events under consideration }and  the temporal decay function $g(\cdot)$  describes how the influence of an earthquake diminishes with time and  follows the Omori-Utsu formula $$g(t) = \frac{p-1}{c}{\left( 1+ t/c\right )^{-p}},$$ with \(t \) being the time lag between the triggering earthquake and triggered earthquake, and \(c>0\) and \(p>1\) are two constants;
\item (c) The magnitudes of triggered events are completely independent of the occurrence times, locations and magnitudes of past events, and the magnitudes themselves are identically and independently distributed, following the Gutenberg-Richter distribution $s(m)=\beta\, \exp[-\beta (m-m_0)]$, which is the probability density function form of the Gutenberg-Richter magnitude-frequency relationship and $\beta = b \ln 10$ is the GR exponent and $b$ is the b-value
often close to $1$.
\end{description}

The mean temporal rate  of this model  is
\begin{equation}\label{eq:etas.mean}
\bar{\lambda}_g:=\ept[\lambda_g(t)] =\frac{\mu }{1-\int_{m_0}^\infty \kappa(m)\,s(m)\, \dd m},
\end{equation}
where  the expected number of offspring from an arbitrary event,
\begin{equation}
\rho:=\int_{m_0}^\infty \kappa(m)\,s(m)\, \dd m = A \beta \frac{1}{\beta-\alpha},~
\label{trwhqteb1}
\end{equation}
 is also called criticality of the process (e.g., \citealp{zhuang2012corssa}) or branching ratio \citep{helmstetter2002jgr}.
 Equation~(\ref{eq:etas.mean}) holds if and only if $\rho < 1$.
In this case, the process is subcritical, ensuring stability and stationarity.
If $\rho \geq 1$, the process becomes critical or supercritical and the mean rate is no longer finite.

The entropy on $[0,T]$ is
\begin{equation}
\ept [-\log L]= \ept\left[ -\sum_{i=1}^{N[0,T]} \log \lambda_g(t_i)+\int_0^T\lambda_g(t)\, \dd t\right]=\ept\left[ \int_0^T \lambda_g(t)(1-\log \lambda_g(t))\, \dd t\right],
\end{equation}
which gives the mean entropy rate  $\epsilon=\ept[ \lambda_g(\cdot)(1-\log \lambda_g(\cdot))]$.

Consider the following general theorem (\citealp{daley2003ppt}):
\begin{theorem}
Let \(N\) be a stationary regular temporal point process on \([0,T]\) with conditional intensity \(\lambda(t)\).
Partition \([0,T]\) into sub-intervals of equal length \(\Delta\), and let
\(X_i \in \{0,1\}\) indicate whether an event occurs in \(((i-1)\Delta,\, i\Delta]\).
Define
\(p_i := \ept[X_i \mid \mathcal{F}_{(i-1)\Delta-}]\),
\(p_i^* := \ept[p_i]\), and
\(m := \ept[\lambda(\cdot)]\) (the average rate). Here, $\mathcal{F}_{(i-1)\Delta-}$ represents the complete event history of the process up to, but not including, time $(i-1)\Delta$
and $p_i := \mathbb{E}\!\left[X_i \mid \mathcal{F}_{(i-1)\Delta-}\right]$  is the conditional probability of observing at least one event in
$((i-1)\Delta,\, i\Delta]$ given the past history. Then, $p_i^* := \mathbb{E}[p_i]$ is the unconditional mean of $p_i$, i.e., the probability of an event in a bin
averaged over all possible histories. Then
\[
G_{\Delta}
:= \ept\!\left\{\frac{1}{T}\sum_{i=1}^{T/\Delta}
\Big[X_i \log\!\frac{p_i}{p_i^*} + (1-X_i)\log\!\frac{1-p_i}{1-p_i^*}\Big]\right\}
\ \ \uparrow\ \
G := \ept\!\left[\lambda(\cdot)\,\log\!\frac{\lambda(\cdot)}{m}\right]
\quad \text{as } \Delta \downarrow 0.
\]
\end{theorem}

It follows that, among processes with the same average rate \(m\), the homogeneous Poisson process (for which \(\lambda(t)\equiv \bar \lambda\)) maximizes entropy, yielding \(G=0\).
For models with time varying intensity, such as the ETAS model, \(\lambda_g(t)\) deviates from \(\bar \lambda_g\), so \(G=\ept[\lambda_g(\cdot)\log(\lambda_g(\cdot)/\bar \lambda_g)] \ge 0\) (with equality only for Poisson).
Thus, the gap between the Poisson and ETAS entropy rates, quantified by \(G\), characterizes the attainable information gain by the true model. In this case,  $G$ is also defined as the {\em intrinsic predictability rate} of the given ETAS model.

Figure \ref{fig:two} illustrates how the entropy rate varies with the ETAS model parameters. In general, models with a higher branching ratio (Figure  \ref{fig:two}a), denser clustering (lower $c$-value; Figure  \ref{fig:two}b)
and more aftershock-sequence like behavior (higher $\alpha$-value, meaning strong magnitude dependence in triggering; Figure  \ref{fig:two}c) exhibit higher predictability.

\newcommand{\figtwo}{Variation of the ETAS entropy rate with different model parameters.
(a) $\alpha$, $c$, $p$, and the total rate $\bar{\lambda}$ are fixed, and the branching ratio varies.
(b) $\mu$, $A$, $\alpha$, and $p$ are fixed, and $c$ varies (a smaller $c$ corresponds to more temporally concentrated clusters).
(c) $\mu$, $c$, $p$, and the branching ratio are fixed, and $\alpha$ varies (a larger $\alpha$ indicates that aftershocks are more likely to follow large earthquakes). }
\begin{figure}
\centerline{\includegraphics[width=0.99\textwidth]{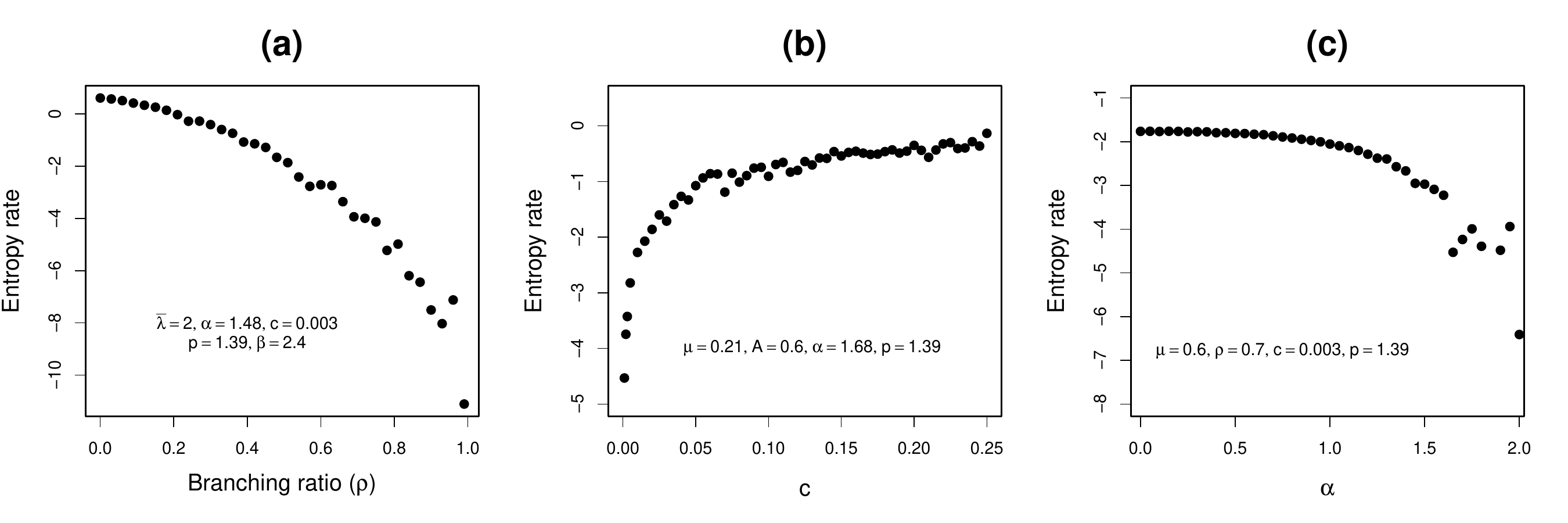} }
\caption{\label{fig:two} \figtwo}
\end{figure}

{It is worthwhile to note that $\alpha=\beta/2$ show a transition point for the entropy rate, as shown in Figure \ref{fig:two}c: F}or $\alpha < \beta/2$, fluctuations are moderate and the entropy rate varies weakly with $\alpha$;
near $\alpha \approx \beta/2$, productivity becomes heavy-tailed, and large bursts begin to dominate;
for $\alpha > \beta/2$, the entropy rate increases sharply because extreme triggering episodes dominate
      the expectation $\mathbb{E}[\lambda(\cdot) \log \lambda(\cdot)]$.
{The above transition can be explained in the following way. In the ETAS model, the productivity associated with an event of magnitude $M$ is
$
K=\kappa(M)=A\exp[\alpha(M-m_0)].
$
Since earthquake magnitudes follow the Gutenberg--Richter distribution,
$
f_M(m)=\beta e^{-\beta(m-m_0)},
$
the productivity is a Pareto random variable:}
\begin{equation}
\Pr(K>x)
=\Pr\left(M>m_0+\frac{1}{\alpha}\log\frac{x}{A}\right) \
=\exp\left[-\frac{\beta}{\alpha}\log\frac{x}{A}\right]
=\left(\frac{x}{A}\right)^{-\beta/\alpha},
\qquad x\ge A.
\end{equation}
{Consequently, the $q$th moment of $K$ is finite if and only if $q\alpha<\beta$. In particular, the mean productivity is finite for $\alpha<\beta$, whereas its second moment is finite only for $\alpha<\beta/2$. Thus, $\alpha=\beta/2$ marks a second-moment transition in the branching structure. Below this threshold, productivity fluctuations remain moderate. Near the transition, the variance of productivity and cluster sizes diverges, whereas above it, rare large-magnitude events dominate the triggering process and produce increasingly bursty seismic activity.}

\section{Wrong model specification}

In the discussion above, we analyzed the performance obtained when the forecasting model coincides with the true model,
as well as when it is misspecified (i.e., when $p^* \neq p$ in the Bernoulli case).
In practice, the true data-generating mechanism is almost never known.
As a result, the forecasting performance of any given model is governed by two distinct factors:
(i) the true underlying process, which determines the intrinsic predictability of the observations, and
(ii) the forecasting model itself, whose structure determines how much of this intrinsic predictability it is able to capture.
In information-theoretic terms, the first factor fixes the irreducible uncertainty (the Shannon entropy of the true process), while the second controls the additional loss due to model misspecification.

Consider the entropy inequality,
\begin{equation}
H(p) := -\int_{\Omega} p(x)\log p(x)\,\mathrm{d}x  = D_{\mathrm{KL}}(p\|p)
\leq -\int_{\Omega} p(x)\log p^{*}(x)\,\mathrm{d}x = D_{\mathrm{KL}}(p\|p^*)
\end{equation}
which holds for any pair of probability density functions $p$ and $p^{*}$.

If we regard $p$ as the true model and $p^{*}$ as the forecasting model, this inequality implies that, in expectation, the forecasting performance of any model cannot exceed that of the true model. However, it remains unclear whether the forecasting performance of a model, quantified by the likelihood ratio against a model of complete randomness, is necessarily bounded by the intrinsic predictability capacity of the forecasting model itself. The answer is negative, since we cannot guarantee that
\begin{equation*}
\int_{\Omega} p^{*}(x)\log p^{*}(x)\,\mathrm{d}x
\geq \int_{\Omega} p(x)\log p^{*}(x)\,\mathrm{d}x.
\end{equation*}
Figure~\ref{fig:one}(c) shows that the expected log-loss $\mathcal{L}(p,p^*)$ is jointly controlled by the true process $p$ and the forecasting model $p^*$
as explained above.
{Importantly, because $H(p)$ for $p=1/4$ is substantially smaller than the maximum entropy $H(1/2)$, the intrinsic predictability of the biased coin is large enough that even moderately misspecified forecasts with $p^*$ between $1/4$ and $1/2$ can still yield a log-loss smaller than that of a completely random model.
{In other words,} the gain in intrinsic predictability of the true system can partially compensate for the degradation caused by imperfect forecasting. This implies,}
Even with $p$ fixed, $\mathcal{L}(p,p^*)$ can be smaller or larger than $H(p^*)$ depending on $p^*$.
This lack of a universal ordering motivates distinguishing models whose realized forecasting performance (against a randomness baseline) appears to exceed, match, or fall short of what one would infer from their own intrinsic uncertainty.

This observation motivates a classification of forecasting models into three categories: \emph{over-performing models},
whose forecasting performance exceeds their own intrinsic predictability, \emph{on-intrinsic models}, whose forecasting
performance exactly the same as their intrinsic predictability,  and \emph{under-performing models}, whose forecasting
performance falls below their intrinsic predictability (see Fig.~\ref{fig:twop}(a) to (c)).

{Under a ``perfect-model'' assumption, a model should not predict real data better than data generated by itself.} When a model outperforms its own synthetic forecasts on real data, it indicates that the real system {is more {predictable} and }contains additional structure or constraints not {fully captured by the forecasting}  model.
This phenomenon typically arises when the model assumes more randomness than the real system actually exhibits. For example, {when} real data may contain hidden regime persistence or structural constraints absent from synthetic simulations, {a lower order stochastic model for forecasting generates ``too random'' samples in the simulation.} In seismology, ETAS-type models may forecast real catalogs better than their own simulations because real seismicity is constrained by fault geometry and long-lived stress heterogeneity not fully represented in homogeneous simulations. Similar effects occur in climate indices and neural systems, where slow components or network architecture confine the system to subsets of states, increasing predictability beyond what the stochastic model assumes (see Fig.~\ref{fig:twop}(c)).

In all these cases, apparent over-performance does not imply that the model is ``too good.'' Rather, it signals that the real system is more organized and less noisy than assumed. The model's internal randomness inflates synthetic uncertainty relative to reality. Such cases therefore indicate unused predictability in the system and point to missing structure?hidden variables, constraints, regimes, or couplings?that remains to be identified.

\newcommand{\figtwop}{(a) Illustration of intrinsic, overperforming, and underperforming forecasting models and their relationships with the expected negative log-likelihood, entropy, and predictability.
(b) Filled contour plot of the difference between the expected forecasting negative log-likelihood and the information entropy of the forecast model for a Bernoulli trial system as a function of the true probability $p$ and the forecast probability $p^*$. The warm-colored regions correspond to over-performing zones, where the forecast achieves a lower expected negative log-likelihood than implied by its own entropy.
(c) Comparison of the model entropies for ETAS models with a fixed average rate but different branching ratios (all other parameters being fixed), and the corresponding negative log-likelihoods obtained when forecasting a synthetic catalog generated with branching ratio $\rho=0.81$.  {The solid black circles show the entropy rate of the ETAS model as a function of $\rho$. The open red circles show the negative log-likelihood of ETAS forecast models with varying $\rho$ but a fixed total rate $\bar{\lambda}=2.0$, evaluated on a synthetic dataset generated from an ETAS model with the true branching ratio $\rho=0.81$.}}
%Panel (a) shows the entropy of Bernoulli trials as a function of the success probability $p$.
%Panel (b) displays the expected negative log-likelihood for a true system with $p=1/4$, plotted as a function of the forecast probability $p^*$.
%Panel (c) compares the true entropy (red curve) with the expected log-loss under varying forecast models (black curve).}
\begin{figure}
%\begin{tabular}{ccc}
%(a)& (b)& (c) \\
%{\includegraphics[trim=30mm 32mm 192mm 40mm,clip,width=0.29\textwidth]{mcs2/slide5.jpg}} &
%{\includegraphics[trim=600 117 30 120,clip,width=0.29\textwidth]{mcs2/slide6.jpg}} &
%{\includegraphics[trim=240 39 55 0,clip,width=0.3\textwidth]{mcs2/slide8.jpg}}
%\end{tabular}
%\centerline{\includegraphics[width=0.995\textwidth]{fig1.pdf}}
\begin{center}
\begin{tabular}{c}
\bf (a)\\
\includegraphics[trim=420 35 40 30,clip,width=0.400\textwidth]{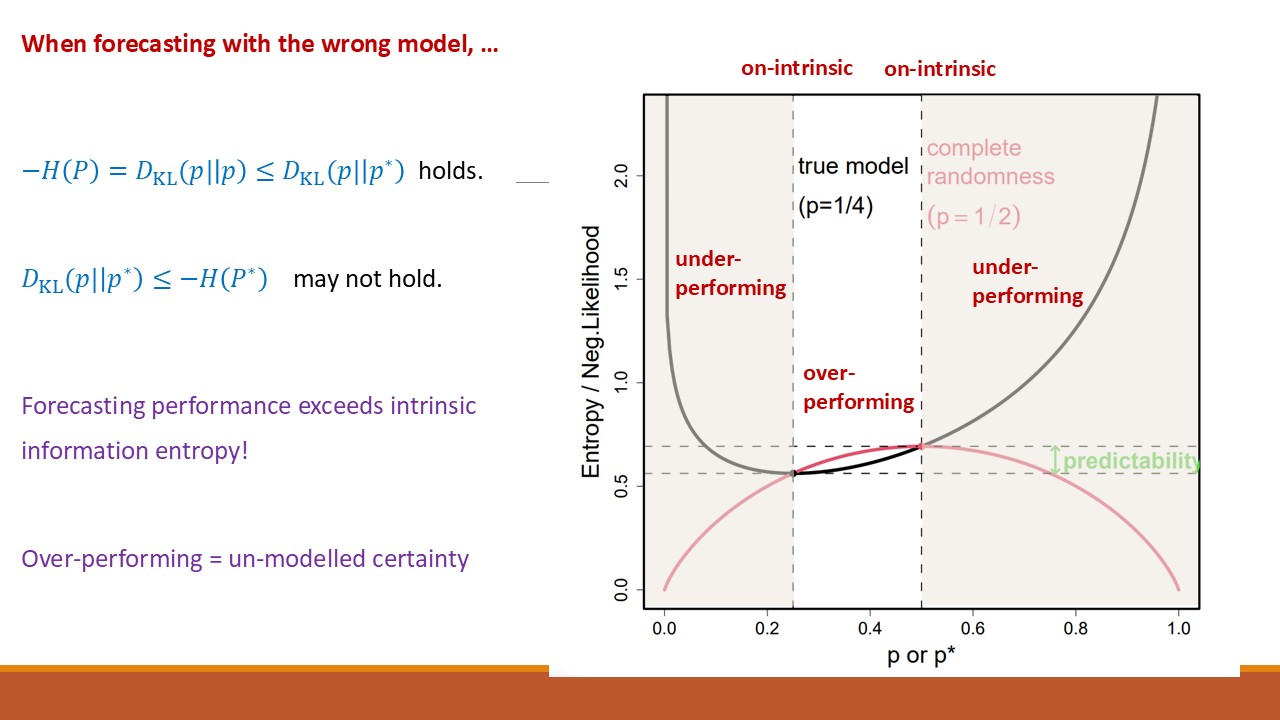}\\ \\ \bf (b) \\
{\includegraphics[trim=-20 0 30 0,clip,width=0.400\textwidth]{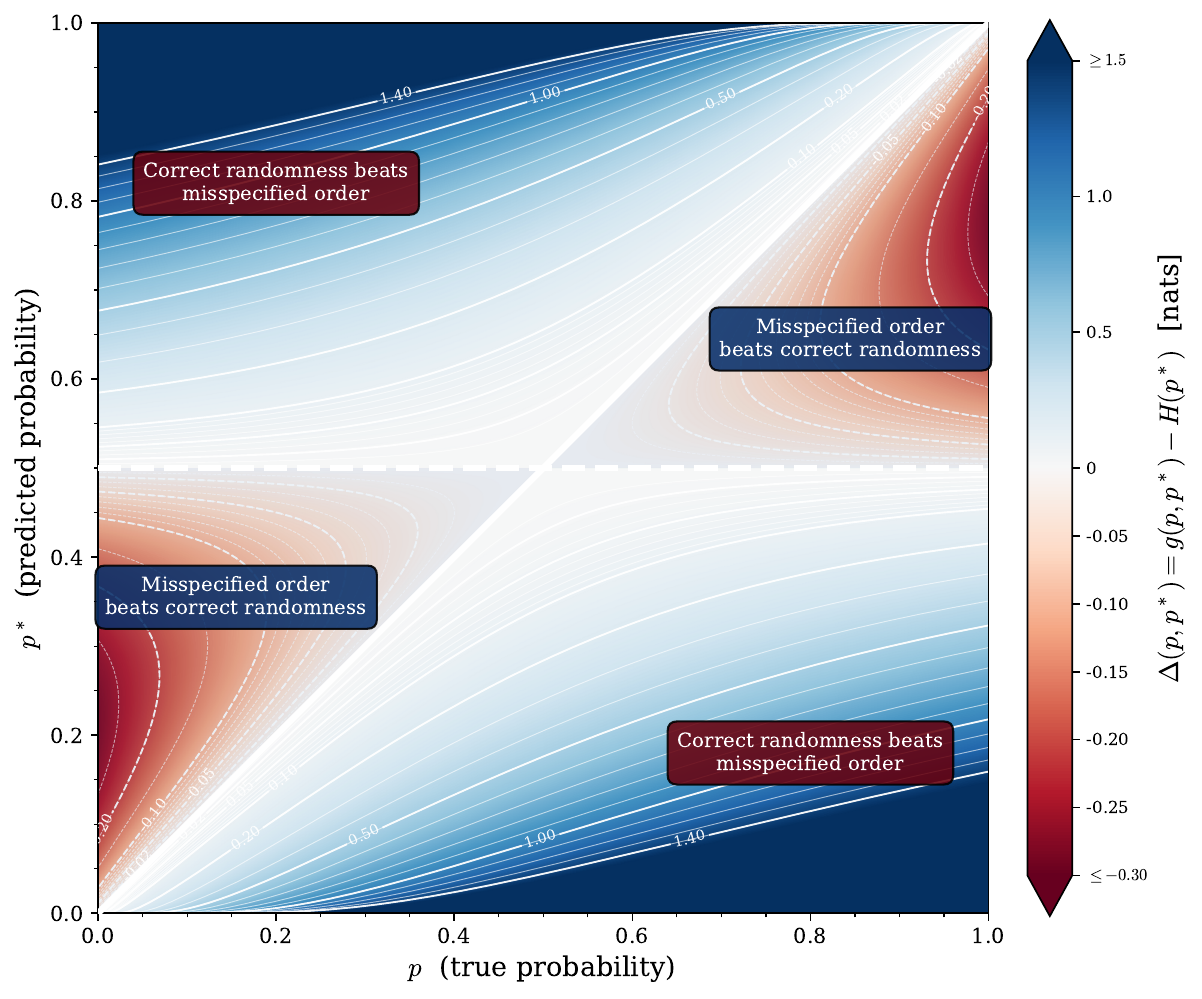}} \\\\\bf (c) \\
\vspace{-0.1cm}\includegraphics[trim=0 20 880 55,clip,width=0.4000\textwidth]{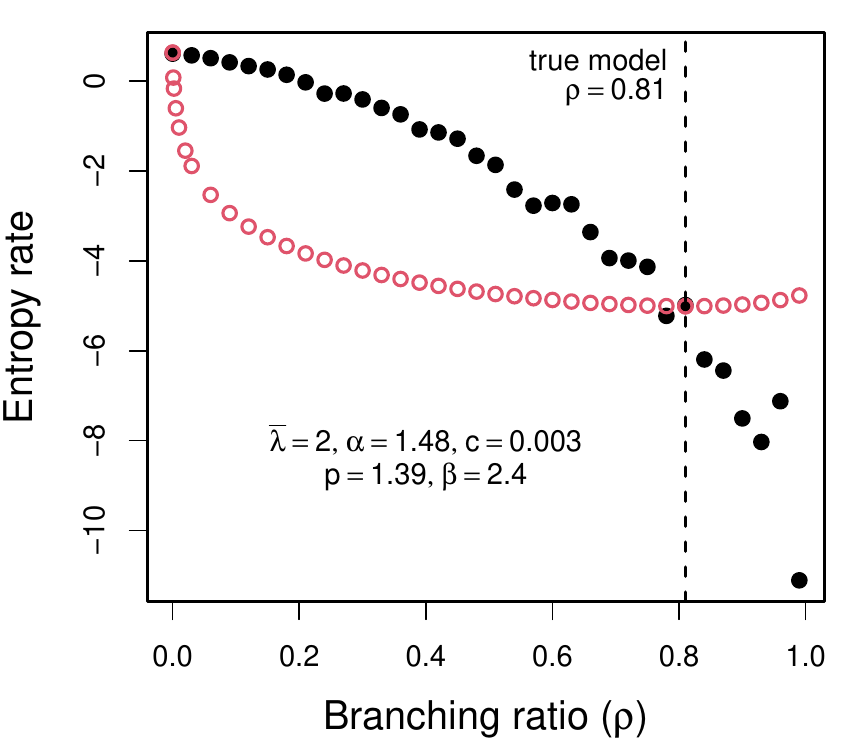}
\end{tabular}
\end{center}
\caption{\label{fig:twop}\figtwop}
\end{figure}

\section{Using a benchmark model}

{At first glance, defining predictability relative to complete randomness may appear inconsistent with recent work advocating comparison against the strongest existing forecasting model \citep{zhang2024jgr,zhang2025esrev,zhang2025gji}. In fact, these approaches address different questions. Entropy reduction relative to a fully random reference quantifies the intrinsic predictability of the earthquake process itself, which provides an upper bound on the structured information available in nature. By contrast, comparing a new model against the best available incumbent measures incremental skill, that is, how much additional structure the new model extracts beyond what is already captured.}

{In practical earthquake forecasting, ETAS-type models currently serve as the strongest established benchmark and are therefore treated as the incumbent reference model. Let $\mathcal{L}(\mathcal{M})$ denote the expected log-loss (cross-entropy) of a forecasting model $\mathcal{M}$ evaluated under the true data-generating process, i.e.,}
\[
{\mathcal{L}(\mathcal{M}) := -\mathbb{E}_{\mathrm{true}} \big[\log p_{\mathcal{M}}(X )\big],}
\]
{where $p_{\mathcal{M}}$ is the {probability density function} of model $\mathcal{M}$. We then define}
\[
{\mathcal{L}(\mathrm{ETAS}) := \mathcal{L}(\mathcal{M}_{\mathrm{ETAS}}),
\qquad
\mathcal{L}(\mathrm{new}) := \mathcal{L}(\mathcal{M}_{\mathrm{new}}),}
\]
{where $\mathcal{M}_{\mathrm{ETAS}}$ denotes the incumbent ETAS benchmark and $\mathcal{M}_{\mathrm{new}}$ the proposed forecasting model. The quantity
$\mathcal{L}(\mathrm{ETAS}) - \mathcal{L}(\mathrm{new})$
then measures the incremental information gain of the new model relative to the strongest existing competitor.}

{These two perspectives are complementary rather than contradictory. The entropy framework defines the total predictability budget of the system, while model-to-model information gain quantifies progress toward exploiting that budget. Formally, letting $H_{\mathrm{rand}}$ denote the entropy of the fully random reference model and $H_{\mathrm{true}}$ the entropy of the true data-generating process, the intrinsic predictability admits the exact decomposition}
\begin{equation}\nonumber
{H_{\mathrm{rand}} - H_{\mathrm{true}}
=
\underbrace{\big(H_{\mathrm{rand}} - \mathcal{L}(\mathrm{ETAS})\big)}_{\text{predictability captured by ETAS}}
+
\underbrace{\big(\mathcal{L}(\mathrm{ETAS}) - \mathcal{L}(\mathrm{new})\big)}_{\text{additional gain of new model}}
+
\underbrace{\big(\mathcal{L}(\mathrm{new}) - H_{\mathrm{true}}\big)}_{\text{remaining gap to optimality}}.}
\end{equation}

{The first term represents the portion of intrinsic predictability already captured by the incumbent ETAS model. The second term quantifies the additional structure extracted by the new model beyond that benchmark. The third term measures the residual gap between the new model and the theoretical optimum (the true process). }

{In this hierarchy, entropy reduction from complete randomness defines the total available predictability, while improvements over the strongest existing model track methodological progress toward realizing that potential. The two viewpoints therefore operate at different levels, one characterizing the physical limits of predictability, the other assessing advances in modeling performance.}  {In the present manuscript, we focus on $H_{\mathrm{rand}} - H_{\mathrm{true}}$, while authors comparing new models to best incumbent models focus on
$\mathcal{L}(\mathrm{ETAS}) - \mathcal{L}(\mathrm{new})$.}

\section{What have we achieved in earthquake predictability?}

\newcommand{\figthree}{Authors' schematic and deliberately conservative assessment of the current state of understanding of earthquake predictability along three dimensions: (a) time, (b) space, and (c) magnitude. }
\begin{figure}
\begin{center}
\begin{tabular}{c}
{\Large\bf{(a)}}\\
\includegraphics[trim=120 102 138 240,clip,width=0.42\textwidth]{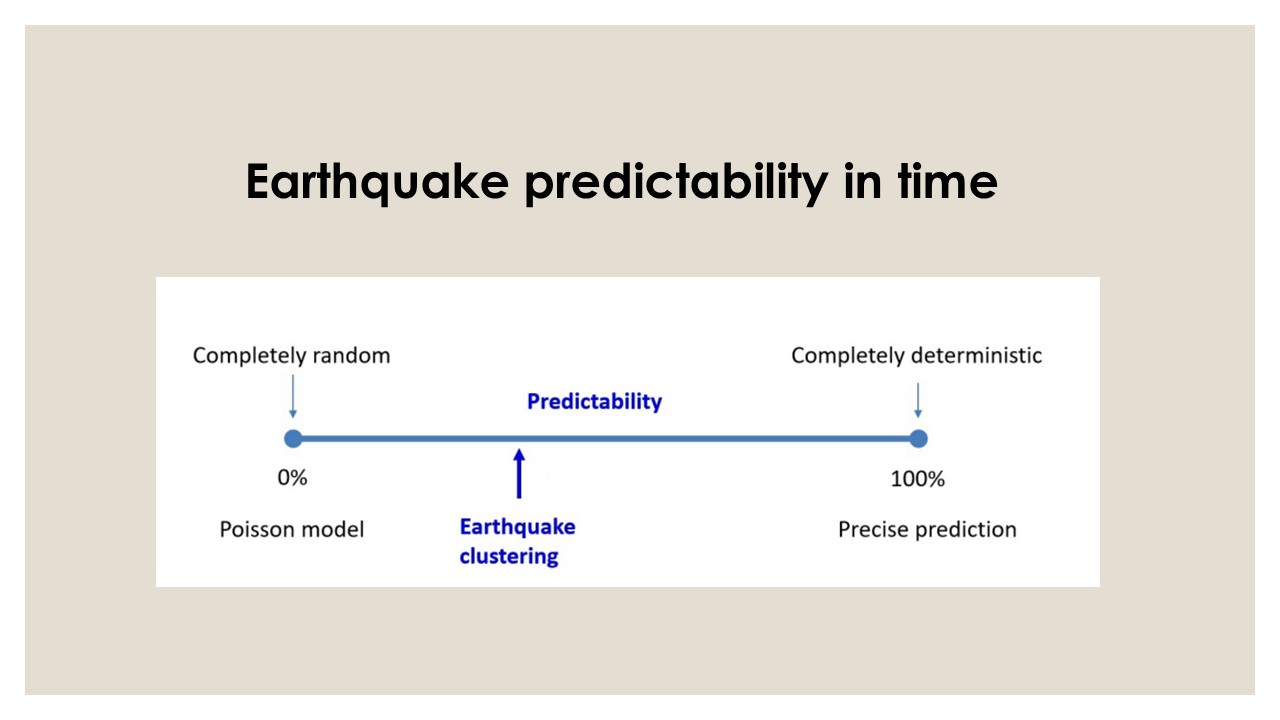}\\\\
{\Large\bf{(b)}}\\
\includegraphics[trim=55 142 60 155,clip,width=0.42\textwidth]{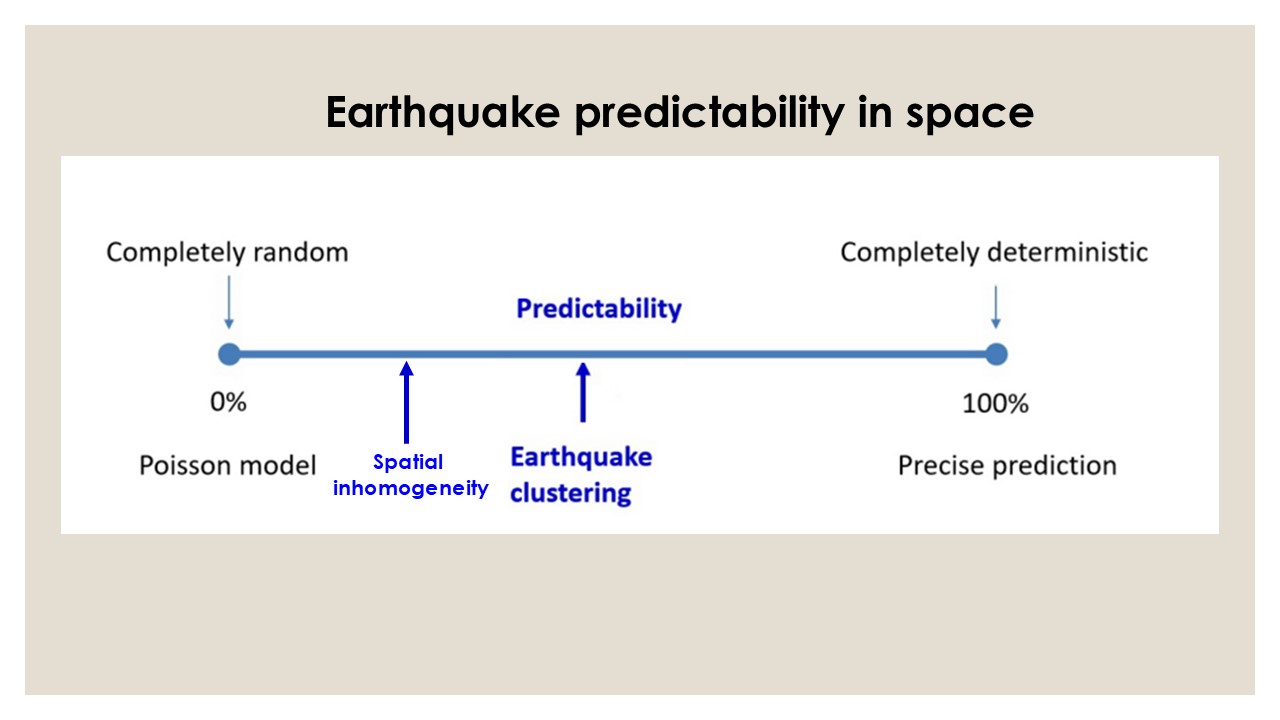}\\\\
{\Large\bf{(c)}}\\
\includegraphics[trim=55 142 60 165,clip,width=0.42\textwidth]{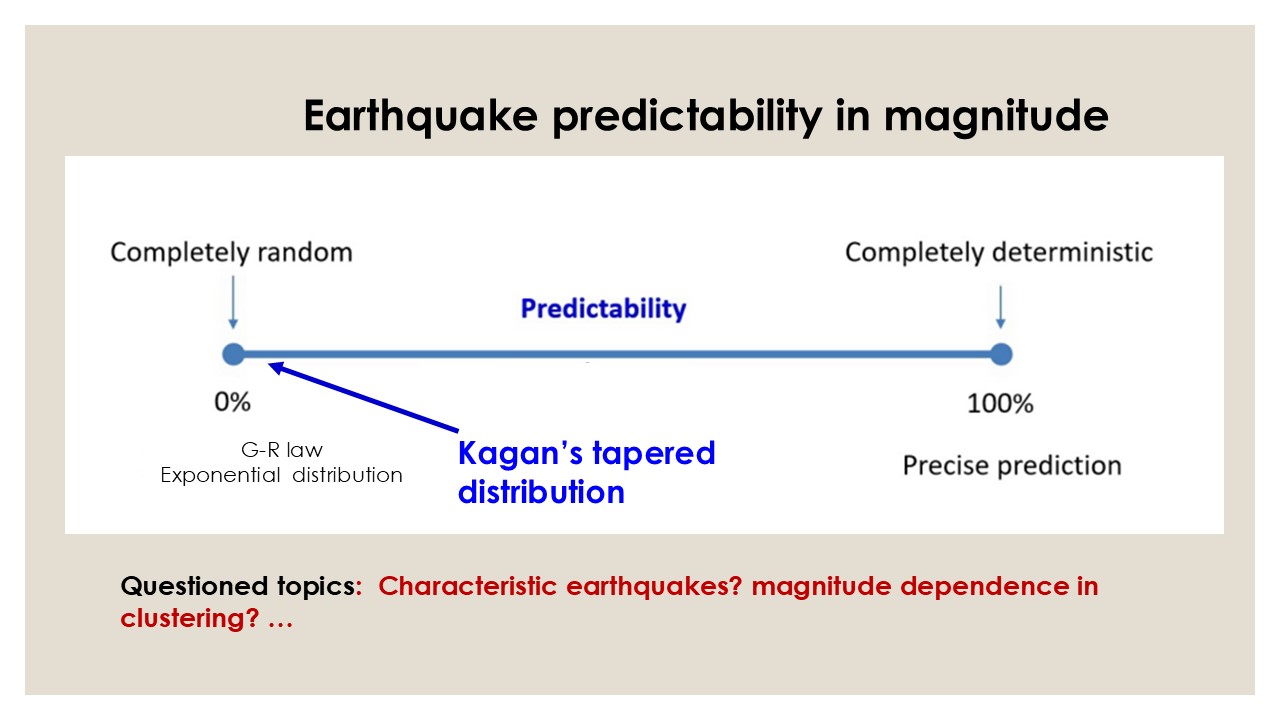}
\end{tabular}
\end{center}
\caption{\label{fig:three}\figthree}
\end{figure}

\subsection{Decomposing earthquake predictability}

Figure~\ref{fig:three} presents a deliberately conservative schematic summary of our current understanding of earthquake predictability in time, space, and magnitude.

Along the temporal axis, complete randomness is represented by a stationary Poisson process, whereas complete determinism would correspond to prediction with 100\% precision. In practice, the ETAS model has become the de facto reference model or null hypothesis for testing alternative forecasting approaches (see, e.g., \citealp{huang2016pageoph,zhuang2021epj,zhuang2023emg,mizrahi2024rg,zhang2025gji,zhang2025esrev}). This widespread adoption implicitly reflects the view that spatiotemporal clustering, particularly aftershock triggering, constitutes the dominant and most robustly established predictable temporal component of seismicity.

Along the spatial axis, complete randomness again corresponds to a homogeneous Poisson process. Beyond the spatial clustering effects partially captured by ETAS, an additional and significant source of predictability arises from spatial inhomogeneity in the background seismicity rate. This inhomogeneity reflects the concentration of earthquakes along active faults and tectonic boundaries, as well as the influence of geological structure, lithology, and thermal variations within the crust.

In the RELM and CSEP experiments, the highest-performing models are almost always variants of ETAS-type models that combine spatiotemporal triggering with spatially variable background rates (e.g. \citealp{NandanOuildi17}). Although various non-seismic precursors have been proposed, their predictive skill has not yet been consistently validated or incorporated into operational forecasting practice.

Along the magnitude axis, complete randomness should be understood as the assumption that the joint distribution of earthquake magnitudes factorizes into the product of individual (marginal) distributions. In other words, magnitudes are assumed to be independent. A further simplification, quite well supported empirically,
is that magnitudes are approximately identically distributed. Together with independence, this corresponds to the simplest form of pure randomness:
independently identically distributed (i.i.d.).
The common distribution of earthquake magnitudes happens to be well-described by an exponential law, called the Gutenberg-Richter (GR) magnitude frequency relation.

In contrast to temporal and spatial predictability, progress in establishing statistically significant magnitude predictability has been limited. The most widely accepted refinement is that a tapered exponential distribution (equivalently, a tapered Pareto distribution for seismic moments) provides a more accurate description of magnitudes at the largest scales, reflecting the finite upper bound imposed by tectonic dimensions (e.g., \citealp{kagan2001jap}). \citet{sornette1999bssa} showed that the tapered distribution arises as the maximum-entropy distribution under the physically motivated constraint that the total seismic moment released per unit time remains finite, reflecting the finite energy budget of a finite earth{, and suggest} a weak form of collective magnitude interdependence.

\subsection{More on earthquake magnitude predictability}
{As mentioned in Section 5.1, beyond the tapered refinement of the marginal distribution, robust and operationally exploitable magnitude correlations have not yet been firmly established. While some studies have reported indications of magnitude dependence, the evidence remains limited and sometimes debated, and no consensus model has emerged that demonstrably improves forecasting performance. The existence and practical relevance of magnitude correlations therefore remain open questions under active investigation.} In the following, we summarize several recent research topics on earthquake magnitude predictability. We do not include these studies in Figure~\ref{fig:three}, as they remain under active debate among researchers; accordingly, we adopt the most conservative viewpoint in the figure.

\paragraph{Characteristic earthquakes} The characteristic earthquake model proposes that individual fault segments tend to rupture in large earthquakes of similar size at quasi-periodic intervals. This view originated from paleoseismic observations along major faults such as the San Andreas and Wasatch Faults, where repeated surface-rupturing events of comparable magnitude were interpreted as evidence for segment-controlled rupture behavior (e.g., \citealp{schwartz1984jgr,sieh1981eEqPred,sieh1989jgr}). Under this framework, each fault segment is thought to host a preferred ``characteristic'' magnitude and to cycle through long-term stress accumulation and release in a relatively regular manner.  However, extensive statistical analyses and modern observations have cast doubt on the universality of characteristic behavior. Earthquake catalogs generally adhere to the Gutenberg?Richter magnitude?frequency relationship without showing the expected peak at a characteristic magnitude (\citealp{kagan1991gji,kagan1993bssa,kagan1997bssa}). Recurrence intervals inferred from paleoseismic records also display significant variability rather than quasi-periodicity (\citealp{grant1994jgr}). Moreover, the failure of the Parkfield prediction experiment?once regarded as a prime example of quasi-periodicity?highlighted the limitations of the model (\citealp{jackson2006bssa}). Advances in dynamic rupture modeling and regional hazard frameworks such as UCERF3 have further supported a transition toward Gutenberg?Richter scaling, renewal models, and physics-based approaches that better capture the complexity and variability of fault behavior (\citealp{field2014bssa}). As a result, the characteristic earthquake model is now considered an oversimplification rather than a general description of fault behavior.

\paragraph{Dragon-king theory} The dragon-king  theory has been proposed as an alternative to both strict Gutenberg-Richter (GR) scale invariance and the characteristic earthquake hypothesis (\citealp{li2025rgpp,li2026cses}).
Whereas the characteristic earthquake model attributes large deviations from the GR law to recurrent, fault-specific ruptures of characteristic size, the dragon-king (DK) hypothesis argues that some very large to great earthquakes transcend individual fault segments and arise from distinct amplification dynamics, making them genuine statistical outliers. In this view, extreme events are not simply the upper tail of a self-similar distribution, nor merely fault-specific characteristic ruptures, but belong to a different statistical and dynamical regime.
{Recent work has developed data-driven procedures to test the DK hypothesis, combining objective spatial clustering with high-power sequential outlier detection to identify anomalous tail events relatie to a GR null model (\citealp{sornette2025joas}). Applications to sequences such as Haicheng (1975) and Tangshan (1976) provide suggestive, though not yet definitive, evidence that some mainshocks may exhibit dragon-king signatures \citep{li2026cses}.}
Numerical simulations of rupture dynamics indicate the coexistence of self-arresting earthquakes and run-away unstable ruptures under different stress and frictional regimes, offering a possible physical basis for departures from strict GR scaling at the largest magnitudes \citep{sornette2026prx}. {If validated, the dragon-king hypothesis would imply that some great earthquakes arise from distinct nucleation mechanisms that may enhance predictability. }However, these ideas remain under active investigation and require prospective validation, ideally through CSEP-style testing frameworks.

\paragraph{Variation of the $b$-value} Spatial variations in the Gutenberg?Richter $b$-value have been widely studied to examine how seismicity differs across fault systems and tectonic environments, providing insight into stress state, material heterogeneity, and fault maturity. High $b$-values are commonly observed in damaged or heterogeneous rock volumes, such as geothermal and volcanic regions, whereas low $b$-values are often associated with high-stress asperities or locked fault segments (e.g., \citealp{wiemer2002bssa, schorlemmer2004jgr}). These spatial patterns have frequently been used to infer seismogenic structures and potential asperity zones.
Spatiotemporal $b$-value analyses further investigate whether temporal changes in the $b$-value reflect evolving stress conditions or rupture processes before or after major earthquakes. Several case studies have reported temporary decreases in the $b$-value prior to large events and increases during aftershock sequences (e.g., \citealp{nanjo2012grl, guila2019nature}).

Despite recent advances in catalog completeness estimation, declustering methods, and the availability of high-resolution seismic catalogs, the robustness of such patterns remains debated. A major concern is the strong sensitivity of $b$-value estimates to catalog completeness and sample size (e.g., \citealp{geffers2022gji}). Many reported fluctuations can be explained by sampling variability or catalog incompleteness rather than genuine physical changes. Using a truncated Gutenberg?Richter law, \citet{marzocchi2020gji} demonstrated that apparent $b$-value biases can arise from correlations with the maximum observed magnitude and from catalog incompleteness. Similarly, \citet{lombardi2024gji} found no significant $b$-value changes in Italy after properly accounting for estimation uncertainty and limited sample precision, highlighting the difficulty of detecting reliable $b$-value variations in practice.

Overall, $b$-value computation requires rigorous treatment of catalog completeness, uncertainty, and statistical significance. Many past claims of meaningful $b$-value variations therefore lack sufficient methodological support, and the predictive value of $b$-value changes must be evaluated more rigorously using prospective testing frameworks such as RELM (\citealp{Schorlemmer-et-al2007,Schorlemmer-et-al2010}) or CSEP  (\citealp{schorlemmer2018srl}) projects.

\paragraph{Foreshocks}
Foreshocks are usually defined retrospectively as events preceding the largest shock within a space--time cluster. Many well-documented earthquake sequences exhibit pronounced foreshock activity, including the 1975 Haicheng earthquake (e.g., \citealp{wang2006bssa}), the 1999 ?zmit earthquake (e.g., \citealp{bouchon2011science}), and the 2011 Tohoku-Oki earthquake (\citealp{kato2012science,yagi2011grl}). Physical models suggest that such foreshocks may reflect preparatory processes prior to dynamic rupture, such as accelerating aseismic slip, damage accumulation, or fluid-related effects (\citealp{dodge1996jgr,ellsworth1995science}). With the advent of high-resolution seismic catalogs and real-time statistical tools, foreshocks have again attracted attention as potential precursors, particularly following the proposal by \citet{guila2019nature} that sharp drops in the Gutenberg--Richter $b$-value might distinguish foreshocks from aftershocks in real time.

Despite these observations, the predictive value of foreshocks remains highly controversial. A fundamental difficulty lies in retrospective bias: earthquakes are labeled as ``foreshocks'' only after the occurrence of a subsequent larger event (\citealp{reasenberg1999jgr}). Large-scale catalog analyses show that only a small fraction of small earthquakes are followed by major events, rendering deterministic prediction based on foreshocks statistically untenable (\citealp{kagan1976pepi,kagan1991gji}). Moreover, earthquake size itself can only be determined after rupture completion, further limiting prospective identification of foreshocks \citep{dvj1976pageoph,dvj1977jiamg,rydelek2006nature,ide2019nature,zhuang2016bssa}.

Within the space--time ETAS framework \citep{ogata1988jasa,ogata1998aism,zhuang2004jgr}, foreshocks, mainshocks, and aftershocks are not distinct categories but arise from the same underlying triggering mechanism \citep{helmstetter2003bjgr,helmstetter2003cjgr}. A central theoretical object is the distribution of the largest descendant magnitude in an ETAS branching tree. Specifically, the probability that an event of a certain magnitude has no descendant exceeding a threshold magnitude governs both (i) the probability that an event becomes a foreshock, i.e., has at least one larger descendant, and (ii) the emergence of the B{\aa}th law as an asymptotic property of the magnitude difference between a parent event and its largest triggered event \citep{helmstetter2003grl,saichev2005pre,zhuang2006pre}. The resulting behavior depends primarily on the productivity?magnitude ratio $\nu=\alpha/\beta$ and the branching ratio $\rho$, with subcritical, critical, and supercritical regimes exhibiting distinct scaling forms \citep{saichev2005pre,zhuang2006pre,dvj2008aap,luo2016bssa}.

A substantial body of observational and simulation studies has shown that foreshock statistics are broadly consistent with ETAS predictions, supporting the view that foreshocks are not fundamentally different from other earthquakes in terms of triggering dynamics \citep{helmstetter2003ajgr,helmstetter2003bjgr,helmstetter2003cjgr,zhuang2006pre,zhuang2008jgr,marzocchi2011grl}. In this interpretation, much of the observed foreshock phenomenology arises naturally from earthquake clustering rather than from special precursory mechanisms.

Claims that ETAS fails to reproduce observed foreshock features \citep{bouchon2013natgeo,ogata2014jgr,lippiello2017pageoph,lippiello2019entropy} appear sensitive to methodological choices. These include fitting ETAS models to quiescent periods, uncertainties in short-time and short-distance parameters, magnitude-threshold effects, and catalog incompleteness immediately following large events \citep{zhuang2008jgr,zhuang2021foreshockbath}. Such issues can systematically bias synthetic foreshock rates and time?distance diagnostics, and simulations with artificially weak clustering should therefore be avoided when testing ETAS-based null hypotheses.

A comprehensive synthesis is provided by \citet{zhuang2021foreshockbath}, who reviewed the relationship between foreshocks and ETAS-type clustering using branching-process theory. He demonstrated that the magnitude distribution of the largest descendant from a given event simultaneously controls foreshock probabilities, with theoretical predictions closely matching observations (see also \citealp{saichev2005pre,zhuang2006pre,luo2016bssa,dvj2008pre}). Overall, these results support a conservative interpretation: most observed foreshock behavior can be understood as natural consequences of earthquake clustering, without invoking distinct preparatory phenomena.The definition of a foreshock is inherently retrospective, since it depends on the later occurrence of a larger event.

From a forecasting perspective, this suggests that monitoring fundamental clustering characteristics?such as the productivity-to-magnitude ratio $\nu=\alpha/\beta$ and the branching ratio $\rho$?may be more informative than attempting to classify foreshocks in real time \citep{zhuang2013pre,zhuang2020epj}. While foreshocks may offer valuable physical insight in specific cases, their general use for operational earthquake prediction remains statistically fragile.

\paragraph{B{\aa}th's law}
Besides foreshocks, another classic empirical regularity associated with earthquake clustering is the B{\aa}th law. The B{\aa}th law states that the magnitude difference between a mainshock and its largest aftershock has a median value of about 1.2 \citep{bath1965tectonophysics,utsu1961geopmag,utsu1970jfs-hkdu}. Because both the B{\aa}th law and foreshocks rely on identifying the ``largest'' event within a cluster, they are intrinsically post hoc concepts. This raises the question of whether they represent independent physical laws or instead emerge as consequences of more fundamental statistical properties of clustered seismicity \citep{dvj1969bssa,lombardi2002aog,console2003jgrb,dvj2006statsei4,dvj2008aap}.

\citet{helmstetter2003grl} first provided a simulation study showing how the ETAS model can re-produce the B{\aa}th law and that, in this approach, the origin of B{\aa}th's law is to be found in the selection procedure used to define mainshocks and aftershocks rather than in any difference in the mechanisms controlling the magnitude of the mainshock
and of the aftershocks. \citet{zhuang2021foreshockbath} also provided a comprehensive synthesis of the relationship between the B{\aa}th law and the ETAS model, clarifying the extent to which ETAS-type clustering can account for the observed B{\aa}th behavior. Using branching-process theory, he showed that the magnitude distribution of the largest descendant triggered by a given event naturally leads to the B{\aa}th law as the asymptotic form of this distribution, with theoretical predictions close to those observed in real seismicity (see also \citealp{saichev2005pre,zhuang2006pre,luo2016bssa,dvj2008pre}).

\paragraph{Magnitude dependence in earthquake clustering}

In the ETAS model formulation (Eq. \ref{eq:etas}), the magnitudes of triggered events  are completely independent of the occurrence times, locations and magnitudes of past events, and the magnitudes themselves are identically and independently distributed, following the Gutenberg-Richter distribution $s(m)=\beta\, \exp[-\beta (m-m_0)]$. This raises the question of whether we should consider a more complicated model that incorporates magnitude dependence in earthquake triggering, with a conditional intensity such as
\begin{equation}\label{eq:gen.etas}
\lambda(t,m)=\mu s_0(m) +\sum_{i:, t_i<t} \kappa(m_i)\, g(t-t_i)\, s(m\mid m_i),
\end{equation}
where $s_0(m)$ and $s(m\mid m_i)$ are the magnitude PDFs for background events and for events triggered by event $i$, respectively.

The first model used in practical seismicity analysis is the BASS (branching aftershock sequence with self-similarity) model, proposed by \citet{holliday2008physicaa}, which has a similar formulation to Eq.~(\ref{eq:gen.etas}).  {In this model, the magnitude of a triggered event follows a truncated Gutenberg--Richter distribution, with the upper truncation determined either by the magnitude of the triggering event or by by a fixed increment above the triggering magnitude. }\citet{zhuang2013pre} showed  that such a BASS model distorts the global Gutenberg--Richter magnitude--frequency relationship.

{A major theoretical development is the self-similar branching model proposed by} \citet{dvj2005aap}{, subsequently generalized and explored by} \citet{saichev2005bpre} and \citet{NandanOuildi19}{. Unlike the standard ETAS model, the Vere-Jones ETAS (V-ETAS) formulation allows the magnitude distribution of triggered events to depend on the magnitude of the triggering event while preserving the global Gutenberg--Richter magnitude distribution. Besides providing a logically consistent self-similar description of earthquake triggering, these models also eliminate the magnitude-threshold (``ultraviolet'') problem of the standard ETAS model} \citep{SornetteWerner2005a,SornetteWerner2005b}, {whereby triggering would otherwise be dominated by arbitrarily small earthquakes.}

{Practically, since the early 2000s, numerous empirical studies have investigated possible magnitude dependence in earthquake triggering, either by examining neighboring earthquake pairs in space and time or by analyzing parent--offspring relationships inferred from stochastic declustering and ETAS-type models. These studies, together with theoretical developments and numerical simulations, have led to differing and sometimes contradictory conclusions (see }\citet{petrillo2022scirep} {for a review).}
{Particularly, pseudo-prospective forecasting experiments have suggested that gV-ETAS models incorporating magnitude-dependent triggering may outperform the standard ETAS formulation} \citep{NandanOuildi19,Nandan2022}{. These results indicate that allowing conditional magnitude dependence may improve the statistical description of earthquake sequences, although the physical origin of this improvement remains under debate.}

{A related but statistically different analysis was performed by} \citet{taroni2024gji}{, who tested serial dependence in the catalog magnitude sequence using runs tests and Pearson correlations between consecutive catalog magnitudes. Taroni (2024) {further argued that the apparent dependence reported in earlier studies is strongly affected by short-term aftershock incompleteness (STAI). After adopting more conservative treatments of catalog completeness, no statistically significant lag-one serial dependence was detected.} It should be noted that this is not the same statistical object as the conditional offspring distribution assumed in the generalized Vere-Jones ETAS model. Taroni's negative result therefore provides evidence against detectable lag-one serial magnitude correlations after conservative treatment of STAI, but it should not be interpreted as a direct refutation of the subtler parent--offspring dependence represented by the gV-ETAS formulation.}
{A direct comparison would require conservative corrections for catalog incompleteness within a parent--offspring ETAS likelihood framework, along the lines of \citet{NandanOuildi19}. However, the two present authors differ in their assessment of whether the treatment adopted in that study is sufficiently rigorous and conservative for the present purpose. A complementary test would be to determine whether the signatures generated in synthetic gV-ETAS catalogs can be reliably detected by Taroni's statistical methodology.}

{Complementary evidence has been obtained using several independent statistical frameworks designed to distinguish genuine magnitude dependence from observational artifacts. Using both a synthetic physical model and the high-resolution 2016 Amatrice--Norcia catalog,} \citet{petrillo2023prl} {reconstructed probabilistic parent--offspring relationships through stochastic declustering and restricted the analysis to events above a conservatively estimated magnitude of completeness. They found no statistically significant parent--offspring magnitude correlation once catalog incompleteness was properly accounted for. Apparent correlations emerged only when smaller events below the completeness threshold were included, suggesting that much of the observed dependence could be explained by short-term aftershock incompleteness (STAI) rather than by intrinsic magnitude dependence.}

{Using a different approach based on magnitude differences between neighboring events,} \citet{lippiello2024prl} {proposed an incompleteness-aware test based on positive magnitude differences. Motivated by the observation that sufficiently large positive magnitude increments are less affected by missing small events, they analyzed relocated California catalogs and reported statistically significant correlations between successive positive magnitude differences that were not reproduced by incomplete ETAS simulations. This result suggests that subtle magnitude dependence may persist after partially accounting for incompleteness. However, because the test is not based on parent--offspring inference or conditional ETAS likelihood analysis, its interpretation still depends on assumptions regarding the detection process, reshuffling procedure, and the adequacy of the null model.}

{More recently,} \citet{zhan2026submitted} {revisited the problem using an information-theoretic framework based on history-dependent neural probability models. Synthetic experiments verified that the method distinguishes genuine magnitude dependence from artifacts caused by STAI and missing small events. Applied to several high-quality California catalogs, it found no statistically significant evidence for intrinsic magnitude dependence once catalog incompleteness was conservatively treated. In contrast,} \citet{berman2026arxiv} {{, who also employed an AI-based approach,} reached a different conclusion.}

{Overall, magnitude dependence remains one of the most important unresolved questions in statistical seismology and earthquake predictability. Theoretical developments, particularly the self-similar Vere--Jones framework, demonstrate that magnitude-dependent triggering can be formulated consistently while preserving the Gutenberg--Richter law and avoiding pathological lower-magnitude behavior. At the same time, empirical evidence remains mixed. Positive indications have been obtained from likelihood--based gV-ETAS forecasting and from analyses of positive magnitude differences, whereas several independent approaches, including probabilistic parent--offspring reconstruction, serial dependence tests, and flexible neural conditional models, have found little or no robust evidence for intrinsic magnitude dependence once catalog incompleteness has been treated conservatively.}

{Taken together, these studies suggest that strong, practically useful magnitude dependence is unlikely, while subtle effects cannot yet be excluded. Future progress will require comparisons performed on common benchmark catalogs under consistent treatments of STAI, routine incompleteness, and branching uncertainty, together with prospective forecasting experiments capable of determining whether magnitude history contains reproducible predictive information beyond that already provided by the Gutenberg--Richter law and the standard ETAS model.}

\section{Improving forecasting skill with high-dimensional observation}

In the above section, we assessed predictability and uncertainty in the earthquake occurrence process using only observations of the process itself. In practice, however, we also monitor other phenomena that may be related to earthquakes, so-called anomalies, which may carry precursory information (e.g., \cite{zhuang2020epj}). We now illustrate how incorporating such auxiliary, high-dimensional observations can improve earthquake predictability, as measured by reductions in Shannon entropy.

The chain rule of Shannon entropy describes the joint uncertainty when $X$ (earthquakes) and $W$ (anomalies) are statistically dependent:
\begin{equation}
  H(X,W) = H(W) + H(X\mid W),
\end{equation}
where
\begin{equation}
  H(X\mid W)
  := \sum_{w\in\mathcal{W}} p(w)\,H(X\mid w)
  = -\sum_{w\in\mathcal{W}}\sum_{x\in\mathcal{X}} p(w)\,p(x\mid w)\,\log p(x\mid w)
\end{equation}
is the uncertainty remaining in $X$ given $W$, where $\mathcal{W}$ and $\mathcal{X}$ are the domain $W$ and $X$, respectively.

Complementing Eq.~(7), we define \emph{predictability} as the reducible uncertainty about a future target given the available information. Formally, for a target variable $X$ (e.g., a seismicity outcome) and an information set $W$ (such as the past catalog, geodetic data, stress proxies, non-seismic precursors, or model state variables), the predictable component is the reduction in uncertainty quantified by the mutual information
\[
I(W;X) = H(X) - H(X \mid W) \ge 0,
\]
which measures how much knowing $W$ decreases the uncertainty of $X$. Equivalently,
\[
I(W;X) = H(X) + H(W) - H(X,W).
\]
When correlations between $W$ and $X$ are strong, the mutual information is not merely a small correction but may represent a leading-order contribution to the total entropy. In this perspective, intrinsic predictability is naturally interpreted as the amount of structured statistical dependence between $W$ and $X$, rather than as an entropy reduction computed under independence assumptions.

Note that mutual information is symmetric, i.e., $I(W;X)=I(X;W)$, and therefore does not by itself encode a forecasting direction (earthquakes $\to$ anomalies versus anomalies $\to$ earthquakes). In forecasting, however, the direction is imposed by the information available at the prediction time. Let $W$ be decomposed into a pre-event (available) part and a post-event (future or response) part, $W=(W_{\mathrm{pre}},W_{\mathrm{post}})$, where $W_{\mathrm{pre}}$ denotes all information available before the target earthquake outcome $X$ is realized (including non-seismic anomalies observed up to the forecast origin), and $W_{\mathrm{post}}$ denotes information observed only after $X$ (e.g., coseismic/postseismic signals or other responses). By the chain rule for mutual information,
\[
I(W;X)=I(W_{\mathrm{pre}};X)+I(W_{\mathrm{post}};X\mid W_{\mathrm{pre}}).
\]
The second term quantifies dependence induced by post-event responses and is not usable for prospective prediction. Accordingly, when discussing predictability improvement from auxiliary observations, we restrict attention to the pre-event component $I(W_{\mathrm{pre}};X)$ (or equivalently the conditional entropy reduction $H(X)-H(X\mid W_{\mathrm{pre}})$), which represents the information about future seismicity that is available before the event occurs.

{This motivates the notion of predictive information, defined as the mutual information between information available before the forecast origin and the future target. In a prospective forecasting problem, the relevant quantity is not the total mutual information $I(W;X)$, because part of $W$ may consist of coseismic or post-event responses unavailable at forecast time. The usable component is}
\begin{equation}
I(W_{\mathrm{pre}};X)=H(X)-H(X\mid W_{\mathrm{pre}}).
\end{equation}
{Forecast improvement should therefore be understood as increasing the amount of this pre-event information that is captured by the forecasting model, or as enlarging $W_{\mathrm{pre}}$ with additional observations $Z$ that provide positive conditional information
$
I(Z;X\mid W_{\mathrm{pre}})>0.
$
In practice, this can be tested by comparing an incumbent model, such as the ETAS model, with an augmented model that includes $Z$. The empirical log-likelihood gain
$
\frac{1}{N} \sum_{k=1}^N \log \frac{p_{\mathrm{aug}}(x_k\mid W_{\mathrm{pre}}, z_k)}{p_{\mathrm{base}}(x_k\mid W_{\mathrm{pre}})}
$
estimates the additional predictive information supplied by $Z$. Thus, high-dimensional observations improve earthquake forecasting only to the extent that
they reduce the conditional entropy of future seismicity beyond what is already achieved by the seismic catalog and incumbent forecasting models.}

\section{Conclusion and discussion}

From an information-theoretic perspective, this article has illustrated how the predictability of the earthquake process can be quantified, and has shown that earthquake forecasting can be regarded as the process of constructing models that capture predictable structure in observed seismicity relative to complete randomness. This viewpoint distinguishes between model consistency, achievable performance bounds, and genuine gains in information content, as summarized below.

\begin{description}
\item (a) Predictability can be quantified by the information entropy, and it lies in the entropy difference from complete randomness. The information entropy measures the degree of uncertainty inherent in a system. Under the framework of information entropy, a process is predictable only to the extent that its entropy is lower than that of a completely random or maximum-entropy reference model. Thus, the ``predictable component'' of seismicity corresponds to the entropy difference between the observed model and the idealized state of full randomness. This perspective provides a unified and quantitative way to compare different forecasting models, assess how much structure exists in earthquake occurrence, and determine how far the system deviates from pure stochastic variability.

\item (b) Improving forecasting is equivalent to  finding more informative model (with lower system entropy rate): limited improvement can be done through model  calibration (for example, better model fitting procedure).  Improving forecasting accuracy ultimately depends on discovering more informative models, that is, models that reduce the entropy rate of the seismicity process by capturing additional structure or physical insights. While calibration techniques such as more accurate parameter estimation, robust smoothing of background rates, or improved declustering can provide incremental gains, they seldom yield substantial improvements if the underlying model fails to represent key physical or statistical features of the system. In other words, refining the fitting procedure can optimize performance within a model class, but cannot overcome limitations inherent to that class. Meaningful advances require expanding or modifying the model structure itself to extract additional information from the data.

\item (c) Assessing the intrinsic predictability of a forecasting model is as important as evaluating the consistency of its predictions. If the model forecasts the observed data better than it forecasts synthetic data generated by itself (i.e., it over-performs), this indicates that the upper limit of the predictability in the data has not yet been attained.

\item (d) Incorporating high-dimensional anomaly observations can reduce uncertainty in earthquake systems by providing additional information beyond the seismic catalog itself. The improvement in predictability is quantified by the reduction in Shannon entropy, or equivalently by the mutual information $I(W_{\mathrm{pre}};X)$ between pre-event observations and future seismicity. Thus, auxiliary observations contribute to forecasting skill to the extent that they encode structured dependence with future earthquake outcomes.

\item (e) Improving magnitude forecasting requires distinguishing clearly between marginal statistics and dependence structure. The Gutenberg-Richter (G-R) law specifies the one-point (marginal) distribution of earthquake magnitudes, but it is agnostic about whether magnitudes are statistically independent or dependent across events. In other words, the G-R model constrains the amplitude distribution of individual magnitudes, while any inter-event magnitude dependence is encoded in the joint distribution.
Thus, enhancing magnitude predictability does not necessarily require abandoning the G-R marginal law; rather, it requires identifying and modeling departures from independence in the multivariate structure of magnitudes, as well as possible physical constraints that shape their joint behavior (e.g., magnitude-dependent triggering, rupture interaction effects, or stress-mediated coupling). Progress along these lines would increase the mutual information between past and future magnitudes, thereby enriching the informational content of seismicity models and potentially improving predictive skill regarding earthquake size.

\end{description}

Taken together, these considerations indicate that earthquake forecasting remains an active and scientifically challenging field in which both progress and fundamental limitations must be carefully assessed. An entropy-based perspective provides a unifying framework for quantifying predictability, comparing different forecasting models, and clarifying how far observed seismicity departs from purely random behavior. Within this framework, it is essential to distinguish between incremental improvements achieved through model calibration and genuine gains in information content, as well as to recognize the intrinsic upper bounds on forecasting performance imposed by the stochastic nature of the system. In this context, magnitude forecasting emerges as a particularly important direction for future research. Particularly, moving beyond the
Gutenberg-Richter law {(exponential distribution) }and the assumption of magnitude independence opens avenues for incorporating additional physical constraints and statistical structure. Continued advances along these lines will not only deepen our understanding of earthquake processes but will also offer the prospect of fundamentally improving the predictive power of seismicity models.

For earthquake forecasting in particular, improving our ability to anticipate large events, even at the cost of reduced performance for small earthquakes or increased false alarms, may yield disproportionate benefits for hazard mitigation and decision-making. Future work should therefore integrate information-theoretic measures of predictability with explicit utility or loss functions that reflect societal, economic, and safety priorities. In this broader framework, predictability becomes not only a property of the physical system, but also a function of how we value outcomes, opening a path toward more decision-relevant and impact-focused seismic forecasting.

%\remark{\em{All models are wrong. Some are reportable. Fewer are useful. The rarest interpret the data.}}

%\remark{\em{-- The real divide lies in whether we feed data, use data, or understand data.}}

\bibliographystyle{agufull04}

\bibliography{my}
\newpage
\newpage
\section*{List of figure caption}
\begin{description}
\item Figure \ref{fig:one}: \figone
\item Figure \ref{fig:two}: \figtwo
\item Figure \ref{fig:twop}: \figtwop
\item Figure \ref{fig:three}: \figthree
%\item Figure \ref{fig:four}: \capffour
%\item Figure \ref{fig:five}: \capffive
%\item Figure \ref{fig:six}: \capfsix
\end{description}
\newpage
\section*{Physical address}

\textbf{Jiancang Zhuang}

Email: zhuangjc@ism.ac.jp

Institute of Statistical Mathematics

10-3 Midori-cho, Tachikawa, Tokyo. 190-8562 Japan
\section*{Declaration of Competing Interests}

The authors acknowledge there are no conflicts of interest recorded.

\end{document}